\documentclass[dvips,eps,pra,tightenlines,twocolumn]{revtex4}

\usepackage{graphicx}
\usepackage{bbm}
\usepackage{amsmath}
\usepackage{amsfonts}
\usepackage{amssymb}
\usepackage{hhline}
\usepackage{theorem}
\theoremstyle{plain}
\newtheorem{theorem}{THEOREM:}
\newtheorem{proof}{PROOF:}

\begin{document}
\title{Bohr's complementarity relation and the violation of the ${\cal CP}
$ symmetry in high energy physics}
\author{Beatrix C. Hiesmayr and Marcus Huber\\
    \small{Faculty of Physics, University of Vienna,
    Boltzmanngasse 5, A-1090 Vienna, Austria}}

\begin{abstract}
We test Bohr's complementary relation, which captures the most
counterintuitive difference of a classical and a quantum world, for
single and bipartite neutral kaons. They present a system that is
naturally interfering, oscillating and decaying. Moreover, kaons
break the  ${\cal CP}$ symmetry (${\cal C}$\dots charge conjugation,
${\cal P}$\dots parity). In detail we discuss the effect of the
${\cal CP}$ violation on Bohr's relation, i.e. the effect on the
``particle--like'' information and the ``wave-like'' information.
Further we show that the quantity that complements the single
partite information for bipartite kaons is indeed concurrence, a
measure of entanglement, strengthening our concept of entanglement.
We find that the defined entanglement measure is independent of
${\cal CP}$ violation while it has been shown that nonlocality is sensitive to ${\cal CP}$ violation.
\end{abstract}
\maketitle

%%%%%%%%%%%%%%%%%%%%%%%%%%%%%%%%%%%%%%%%%%%%%%%%%%%%%%%%%%%%%%%%
%%%%%%%%%%%%%%%%%%%%%%%%%%%%%%%%%%%%%%%%%%%%%%%%%%%%%%%%%%%%%%%%
\section{Introduction}
%%%%%%%%%%%%%%%%%%%%%%%%%%%%%%%%%%%%%%%%%%%%%%%%%%%%%%%%%%%%%%%%
%%%%%%%%%%%%%%%%%%%%%%%%%%%%%%%%%%%%%%%%%%%%%%%%%%%%%%%%%%%%%%%%

Bohr's complementarity principle or the closely related concept of
duality in interferometric or double slit like devices are at the
heart of quantum mechanics. The complementarity principle was
phrased by Niels Bohr in an attempt to express the most fundamental
difference between classical and quantum physics. According to this
principle, and in sharp contrast to classical physics, in quantum
physics we cannot capture all aspects of reality simultaneously, the
information content obtained in one single setup is always limited.
By choosing the setup, e.g. the double slit parameters, and thus the
quantum state under investigation, the predictability, the \textit{a
priori} knowledge on the path taken, is simply calculated
(``particle--like'' information), whereas the contrast of the
interference pattern (``wave--like'' information) is observed by the
experimenter. In the case of a pure state, the sum of the squares of
these two quantities adds up to one, meaning that the whole
available information, particle--like and wave--like, is conserved.

This principle has intensely been investigated both in theory and
experiment mainly for photons, electrons and neutrons propagating
through a double slit or through an interferometer. We here present
a system which seems to be optimal for testing Bohr's complementary
relation: It's a system which is `created'' by Nature already as an
interfering system ---thus there is no need for modeling by any
experimenter--- and are at the realm of an energy scale which are
usually not used to test the fine workings of quantum physics.
Moreover, it has a natural time evolution, oscillation and decay. In
addition the system breaks a symmetry in high energy physics
\cite{CPExperiment}, the symmetry ${\cal CP}$ (${\cal C}$\dots
charge conjugation, ${\cal P}$\dots parity), which is well studied,
but lacks a deep understanding. Such a candidate is the neutral
K--meson, shortly called kaon. One goal of this work is to discuss
the effect of the symmetry violation to Bohr's complementarity
principle.

%However, it is also of interest to study other systems, e.g., which
%are ``created'' by Nature already as an interfering system ---thus
%there is no need for modeling by any experimenter--- and are at the
%realm of an energy scale which are usually not used to test the fine
%workings of quantum physics. Such a candidate is the neutral
%K-meson, shortly called kaon. Furthermore, they posses properties
%such as violating a symmetry in high energy physics not available by
%other systems. One goal of this work is to discuss the effect of the
%symmetry violation to Bohr's complementarity principle.

We also go one step further and discuss an extension of Bohr's idea
to capture the information theoretic content of bipartite systems.
For bipartite qubits \cite{Bergou} it has been shown that the
missing part adding to the single properties of one partner of the
pair is nothing else than concurrence. This is a computable function
of entanglement of formation which is a measure of entanglement. We
apply that to neutral kaons which is not straightforward, because,
though neutral kaons are two--state systems, they have to be handled
as an open quantum system due to their decay property
\cite{BGH4,H2}. We demonstrate that this can be done in a useful and
consistent way and it strengthens the result about concurrence
presented in Ref.~\cite{H2}.

%Neutral kaons are famous in particle physics because of their
%violation of the symmetry ${\cal CP}$ (${\cal C}$\dots charge
%conjugation, ${\cal P}$\dots parity) which was shown in a remarkable
%experiment \cite{CPExperiment} in 1964, for which the Nobel prize
%was given. We analyze in this work how this ${\cal CP}$--violating
%effect influences Bohr's complementarity relation for single and
%bipartite neutral kaons.

The work is organized as follows. We start by introducing Bohr's
quantitative complementarity relation for single and bipartite
qubits, then we give a short introduction into the quantum formalism
of neutral kaons, followed by deriving both relations, single and
bipartite, for neutral kaons and by discussing the effect of the
symmetry violation.

%%%%%%%%%%%%%%%%%%%%%%%%%%%%%%%%%%%%%%%%%%%%%%%%%%%%%%%%%%%%%%%%
%%%%%%%%%%%%%%%%%%%%%%%%%%%%%%%%%%%%%%%%%%%%%%%%%%%%%%%%%%%%%%%%
\section{Bohr's quantitative complementarity relation}
%%%%%%%%%%%%%%%%%%%%%%%%%%%%%%%%%%%%%%%%%%%%%%%%%%%%%%%%%%%%%%%%
%%%%%%%%%%%%%%%%%%%%%%%%%%%%%%%%%%%%%%%%%%%%%%%%%%%%%%%%%%%%%%%%

 The qualitative well-known statement that ``\textit{the observation
of an interference pattern and the acquisition of which--way
information are mutually exclusive}'' has been rephrased to a
quantitative statement first by Greenberger and
Yasin~\cite{GreenbergerYasin} and then refined by
Englert~\cite{Englert}:
\begin{eqnarray}\label{comp}
{\cal P}^2+{\cal V}^2\leq 1\;,
\end{eqnarray}
where the equality is valid for pure and the inequality for mixed
quantum states. ${\cal V}$ is the fringe visibility which quantifies
the sharpness or contrast of the interference pattern (``the
wave--like property''), whereas ${\cal P}$ denotes the path
predictability, i.e., the \textit{a priori} knowledge one can have
on the path taken by the interfering system (``the particle--like
property''). In double slit experiment it is simply defined by
${\cal P}\;=\;|p_I-p_{II}|$, where $p_I$ and $p_{II}$ are the
probabilities for taking each path ($p_I+p_{II}=1)$.

In the following we are not limiting to the double slit scenario,
rather we write our interfering system generally in the
computational basis
\begin{eqnarray}
\tau&=&\frac{1}{2}\left\lbrace
\mathbbm{1}_2+\vec{n}\cdot\vec{\sigma}\right\rbrace
\end{eqnarray}
where the Bloch vector $\vec{n} \in \mathbbm{R}^3$ and $\left|
\vec{n} \right|^2 \leq 1$ and $\sigma$'s denote the Pauli matrices.
Then the predictability and visibility can be expressed by
\begin{eqnarray}
{\cal P}&=&\left|Tr(\sigma_z\,\tau)\right|\,=\,|n_z|\\
{\cal V}&=&\left|Tr(\sigma^+\,\tau)\right|\,=\,|n_x+i\, n_y|\;,
\end{eqnarray}
and clearly Bohr's relation (\ref{comp}) holds. For all pure states,
$\left| \vec{n} \right|^2=1$, no information is lost, for mixed
states, $\left| \vec{n} \right|^2\leq 1$, the loss is due to
ignorance of individual particles, thus is a purely classical loss.

Note that here the ${\cal V}$ is defined as the coherent
superposition of the two orthogonal states, i.e. as coherence, which
for qubits coincides with the visibility defined by the term that
multiplies the interference term. This is not always the case, i.e.
for neutral kaons, only if the Pauli matrices are chosen in the
strangeness basis coherence coincides with visibility (see
Section~\ref{chapcomsingle}).

One can make Bohr's complementary relation always exact by adding
the quantity
\begin{eqnarray}\label{qubitmixedness}
M^2(\tau)=2((Tr\tau)^2-\, Tr(\tau^2))
\end{eqnarray}
to the single particle property ${\cal S}^2={\cal P}^2+{\cal V}^2$
\begin{eqnarray}\label{compmixed}
{\cal S}^2(\tau)+ M^2(\tau)= 1
\end{eqnarray}
for all states (pure: $M(\tau)=0$). $M(\tau)$ measures the mixedness
or linear entropy which equals in this case the uncertainty of
individual particles under investigation, clearly a ``classical''
uncertainty.

%%%%%%%%%%%%%%%%%%%%%%%%%%%%%%%%%%%%%%%%%%%%%%%%%%%%%%%%%%%%%%%%
%%%%%%%%%%%%%%%%%%%%%%%%%%%%%%%%%%%%%%%%%%%%%%%%%%%%%%%%%%%%%%%%
\section{Complementarity of bipartite qubits and concurrence}
%%%%%%%%%%%%%%%%%%%%%%%%%%%%%%%%%%%%%%%%%%%%%%%%%%%%%%%%%%%%%%%%
%%%%%%%%%%%%%%%%%%%%%%%%%%%%%%%%%%%%%%%%%%%%%%%%%%%%%%%%%%%%%%%%

Let us now proceed to bipartite qubits in a state $\rho$.
Considering one subsystem, clearly we have ${\cal S}_k^2={\cal
P}_k^2+{\cal V}_k^2\leq 1$ where $k$ denotes the chosen subsystem
(${\cal P}_k\equiv{\cal P}(\rho_k),{\cal V}_k={\cal V}(\rho_k)$ and
$\rho_k=Tr_{\neg k}\rho$ is the partial trace over the other subsystem). As Jakob and Bergou \cite{Bergou}
showed the following relation holds
\begin{eqnarray}\label{compbipartite}
\underbrace{{\cal P}_k^2+{\cal V}_k^2}_{\textrm{one--particle
property}\;{\cal S}}+\underbrace{{\cal C}^2}_{\textrm{two--particle
property}}\leq 1
\end{eqnarray}
where the equality sign is valid for all pure bipartite states. Thus
the missing information adding to the single qubit information Alice
or Bob is possessing, is a two--particle property and, surprisingly,
nothing else than the concurrence, ${\cal C}\equiv{\cal C}(\rho)$,
introduced by Hill and Wootters~\cite{Wootters}.

To compute concurrence one defines the flipped matrix
$\tilde\rho=(\sigma_y\otimes\sigma_y)\rho^*(\sigma_y\otimes\sigma_y)$
where $\sigma_y$ is the $y$--Pauli matrix and the complex
conjugation is taken in computational basis. The
concurrence is then given by the formula
$\mathcal{C}=max\{0,\lambda_1-\lambda_2-\lambda_3-\lambda_4\}$ where
the $\lambda_i$'s are the square roots of the eigenvalues, in
decreasing order, of the matrix $\rho\tilde\rho$.

Concurrence is a computable function of entanglement of formation,
which is a measure of entanglement. It is nondecreasing under local
operation and classical communication (LOCC) and does only depend on
the density matrix and is consequently independent on any local
basis. Entanglement of formation is defined by
$\mathcal{E}o\mathcal{F}(\rho)=min_i \sum_i p_i S(Tr_l(
|\psi_i\rangle\langle\psi_i|))$ where $S$ is the von Neumann entropy
$S(\rho)=-\rho\ln\rho$, the trace is taken over one subsystem (left
or right) and $\psi_i$ are the pure state decompositions of
$\rho=\sum_i p_i\,|\psi_i\rangle\langle\psi_i|$ with $0\leq p_i\leq
1, \sum p_i=1$. There are certain criteria a measure should fulfill
(see e.g. Ref.~\cite{Bruss}) among them entanglement of formation is
a good one. The outstanding problem of its additivity has recently
been proven \cite{Paz-SilvaReina}. Concurrence is a computable
function of this measure in the case of a bipartite system with two
degrees of freedom, but it lacks additivity. The more interesting is
it that concurrence is the missing two--particle information adding
to the one--particle information, i.e. complementing the single
particle information the experimenter Alice or Bob can obtain. Of
course, concurrence has no operational meaning, but as shown in
Ref.~\cite{Melo} for a restricted class of states quantum
non--demolition tests of bipartite complementarity can be realized.

Clearly, for mixed states the complementarity relation is exact by
exchanging the concurrence by $M(\rho_k)$. This mixedness obviously
contains the entanglement property as well as classical uncertainty
over individual particles and to separate these has so far only been
achieved for bipartite qubits. However, for
multi--particle or higher bipartite systems ideas to extend Bohr's
principle have been discussed, Ref. ~\cite{Tessier}.

Of course it remains an open question, whether the complementarity
relation (\ref{compbipartite}) is a universal physical feature of
all quantum systems and in what respect. Therefore we move on to the
neutral kaon system which is a two--state system, i.e. only two
degrees of freedom can be measured, however, by decaying and
breaking the ${\cal CP}$ symmetry, differs from ``normal'' qubits.

%%%%%%%%%%%%%%%%%%%%%%%%%%%%%%%%%%%%%%%%%%%%%%%%%%%%%%%%%%%%%%%%
%%%%%%%%%%%%%%%%%%%%%%%%%%%%%%%%%%%%%%%%%%%%%%%%%%%%%%%%%%%%%%%%
\section{Neutral kaons and their time
evolution}\label{thetimeevolutionkaon}
%%%%%%%%%%%%%%%%%%%%%%%%%%%%%%%%%%%%%%%%%%%%%%%%%%%%%%%%%%%%%%%%
%%%%%%%%%%%%%%%%%%%%%%%%%%%%%%%%%%%%%%%%%%%%%%%%%%%%%%%%%%%%%%%%

There have been a lot of puzzles about neutral kaons before a correct
description was found. We here introduce only shortly the quantum
mechanical formulation of these particles and work out the
differences to qubits, which we need for analyzing Bohr's
complementarity relation in the next section.

By the quantum number $S$, the strangeness, which is conserved for
strong interaction, we can distinguish between two states, the
particle $K^0$ and the antiparticle $\bar K^0$. Both strangeness
states can decay via weak interaction into the same decay products,
thus enables strangeness oscillation: if a $K^0$ is produced at time
$t=0$, one finds at a certain later time $t$ a $\bar K^0$. Thus
neutral kaons oscillate between their particle and antiparticle
state and have to be handled as a two--state system.

The time evolution is usually described via an effective
Schr\"odinger equation which we write in the Liouville von Neumann
form as
\begin{eqnarray}\label{effectiveSchroedi}
\frac{d}{dt}\, \tau_{ss}&=&-i\, H_{eff}\, \tau_{ss} + i\,
\tau_{ss}\, H_{eff}^\dagger
\end{eqnarray}
where $\tau_{ss}$ is a $2\times 2$ matrix and the Hamiltonian
$H_{eff}$ is non-Hermitian. Using the Wigner-Weisskopf-approximation
the effective Hamilton can be defined to be
$H_{eff}=H-\frac{i}{2}\Gamma$ where the mass matrix $H$ and the
decay matrix $\Gamma$ are both Hermitian and positive. This Wigner-Weisskopf approximation gives the exponential
time evolution of the two diagonal states of $H_{eff}$:
\begin{eqnarray}\label{exptime}
|K_{S}(t)\rangle &=& e^{-i \lambda_{S} t}\;
|K_{S}\rangle\,,\nonumber\\
|K_{L}(t)\rangle &=& e^{-i \lambda_{L} t}\; |K_{L}\rangle\,,
\end{eqnarray}
with $\lambda_{S/L}=m_{S/L}-i \frac{\Gamma_{S/L}}{2}$ where $m_{S/L}$
and $\Gamma_{S/L}$ are the masses and decay constants for the
short/long--lived state $K_{S/L}$. What makes the neutral kaon
systems so attractive for many physical analyzes, as e.g.
considering Bell inequalities (e.g. Refs.~\cite{BH1,BGH3}) or quantum marking
and eraser experiments~\cite{SBGH1,SBGH6}, is the huge factor
between the two decay rates, i.e. $\Gamma_S\approx 600 \Gamma_L$,
and that the strangeness oscillation is $\Delta
m=m_L-m_S\simeq\Gamma_S/2$.

A kaon or an antikaon is a superposition of the two
mass--eigenstates $K_S/K_L$ or the two ${\cal CP}$ eigenstates
denoted by $K_1^0/K_2^0$:
\begin{eqnarray}\label{states}
| K^0\rangle &=&\frac{N}{2p} \{\hphantom{-} |K_S\rangle + |K_L
\rangle\}=\frac{1}{\sqrt{2}} \{
\hphantom{-} |K_1^0 \rangle + |K_2^0\rangle \}\nonumber\\
| \bar K^0\rangle &=& \frac{N}{2q} \{ -|K_S\rangle + |K_L
\rangle\}=\frac{1}{\sqrt{2}} \{ -| K_1^0 \rangle + |K_2^0\rangle
\}\,.\nonumber\\
\end{eqnarray}
The weights $p=1+\varepsilon$, $\,q=1-\varepsilon,\,$ with
$N^2=|p|^2+|q|^2$ contain the complex ${\cal CP}$ \textit{violating
parameter} $\varepsilon$ which is measured to be
$\varepsilon=(2.28\pm0.02)\times10^{-3}e^{i\frac{\pi}{4}}$. It means that
the short--lived K--meson decays dominantly into $K_S\longrightarrow
2 \pi$ with a lifetime $\tau_S=\Gamma^{-1}_S$ and the long--lived
K--meson decays dominantly into $K_L\longrightarrow 3 \pi$ with a
lifetime $\tau_L=\Gamma^{-1}_L$. However, due to $\cal{CP}$
violation we observe a small amount $K_L\longrightarrow 2 \pi$. This
introduces a small, but measurable difference between a world made
of matter and a world made of antimatter.

Considering Eq.~(\ref{exptime}) we notice that the state is not
normalized for $t>0$. Indeed, we are not describing a system, for
$t>0$ a neutral kaon has a surviving and decaying component.
Mathematically, the Hilbert space for single kaon evolving in time
has to be divided into a direct sum, i.e.
$\textbf{H}_{tot}=\textbf{H}_s\bigoplus\textbf{H}_f$ where $s/f$
denotes ``surviving'' and ``decaying'' or ``final'' components.

There exist two approaches to view the kaonic system. One defines
that the state for a short or long lived kaon after it propagates a
certain time $t$ is given by
\begin{eqnarray}
|K_{S/L}(t)\rangle&=&\underbrace{e^{-i \lambda_{S/L} t}
|K_{S/L}\rangle}_{surviving}+\underbrace{|\Omega_{S/L}(t)\rangle}_{decaying}
\end{eqnarray}
where the time evolution of the decaying components is obtained via
\begin{eqnarray}
1&\stackrel{!}{=}&\langle K_{S/L}(t)
|K_{S/L}(t)\rangle=e^{-\Gamma_{S/L} t}+\langle \Omega_{S/L}(t)
|\Omega_{S/L}(t)\rangle\;.\nonumber\\
\end{eqnarray}
Differently stated, everything that is lost due to decay is added to
the decaying components, such that the time evolution of the whole
system is unitary. Mathematically, the Hilbert space is a direct sum
and the $``+"$ sign in the two above equations has to be understood
in this way. The advantage working with this picture is that for
initial pure states one deals only with the wave function formalism,
the disadvantage is the time evolution of the decaying components
cannot explicitly be given.

Another picture for kaons is given by an open quantum approach to
particle decay \cite{BGH4,H2}: As time evolves the kaon interacts
with an environment which causes the decay (in our case the
environment plays the role as the QCD vacuum in quantum field
theory, but has not to be modeled). In particular the time evolution
of neutral kaons is described by a master equation
\cite{Lindblad,GoriniKossakowskiSudarshan}
\begin{eqnarray}\label{masterequation}
\frac{d}{dt} \tau&=&-i [\cal{H},\tau]-\cal{D}[\tau]
\end{eqnarray}
where the dissipator under the assumption of complete positivity and
Markovian dynamics has the well known general form $ {\cal
D}[\tau]=\frac{1}{2}\sum_j ({\cal A}_j^\dagger{\cal
A}_j\tau+\tau{\cal A}_j^\dagger{\cal A}_j-2 {\cal A}_j\tau{\cal
A}_j^\dagger)$. The density matrix $\tau$ lives on
$\textbf{H}_{tot}=\textbf{H}_s\bigoplus\textbf{H}_f$ %where $s/f$
%denotes ``surviving'' and ``decaying'' or ``final'' components,
and has the following decomposition
\begin{equation}\label{rhotot}
\tau=\left(\begin{array}{cc} \tau_{ss}&\tau_{sf}\\
\tau_{sf}^\dagger&\tau_{ff}\end{array}\right)
\end{equation}
where $\tau_{ij}$ with $i,j=s,f$ denote $2\times 2$ matrices. The
Hamiltonian $\cal{H}$ is the Hamiltonian $H$ of the effective
Hamiltonian $H_{eff}$ extended to the total Hilbert space
$\textbf{H}_{tot}$ and $\Gamma$ of $H_{eff}$ defines a Lindblad
operator by $\Gamma=A^\dagger A$, i.e.
\begin{eqnarray*}
{\cal H}=\left(\begin{array}{cc} H&0\\
0&0\end{array}\right)\;,\; {\cal A}=\left(\begin{array}{cc} 0&0\\
A&0\end{array}\right)\quad\textrm{with}\quad A:
\textbf{H}_s\rightarrow \textbf{H}_f\,.
\end{eqnarray*}
Rewriting the master equation for $\tau$, Eq.~(\ref{rhotot}), on
$\textbf{H}_{tot}$
\begin{eqnarray}
\dot{\tau}_{ss}&=&-i[H,\tau_{ss}]-\frac{1}{2}\,\lbrace A^\dagger
A, \tau_{ss}\rbrace\;,\\
\dot{\tau}_{sf}&=&-i H \tau_{sf}-\frac{1}{2}\, A^\dagger
A\, \tau_{sf}\;,\\
\label{rhoff} \dot{\tau}_{ff}&=&A\,\tau_{ss}\,A^\dagger\,,
\end{eqnarray}
we notice that the master equation describes the original effective
Schr\"odinger equation (\ref{effectiveSchroedi}) but with properly
normalized states, Ref.~\cite{BGH4}. By construction the time
evolution of $\tau_{ss}$ is independent of $\tau_{sf}, \tau_{fs}$
and $\tau_{ff}$. Further $\tau_{sf},\tau_{fs}$ completely decouples
from $\tau_{ss}$ and thus can without loss of generality be chosen
to be zero, they are not physical and can never be measured. With
the initial condition $\tau_{ff}(0)=0$ the time evolution is
\textit{solely} determined by $\tau_{ss}$---as expected for a
spontaneous decay process---and formally given by integrating
Eq.~(\ref{rhoff}). It proves that the decay is Markovian and
moreover completely positive.

Explicitly, the time evolution of a neutral kaon is given in the
lifetime basis, $\{K_S, K_L\}$, by ($\tau_{SS}+\tau_{LL}=1$):
\begin{widetext}
\begin{eqnarray}\label{densitysingle}
\tau(t)=\left(\begin{array}{cccc} e^{-\Gamma_S t} \tau_{SS}& e^{-i
\Delta m t-\Gamma t} \tau_{SL}&0&0\\
e^{i \Delta m t-\Gamma t} \tau_{SL}^*&e^{-\Gamma_L t} \tau_{LL}&0&0\\
0&0&(1-e^{-\Gamma_L t}) \tau_{LL}&0\\
0&0&0&(1-e^{-\Gamma_S t}) \tau_{SS}
\end{array}\right)\,.
\end{eqnarray}
\end{widetext}
%%%%%%%%%%%%%%%%%%%%%%%%%%%%%%%%%%%%%%%%%%%%%%%%%%%%%%%%%%%%%%%%
%%%%%%%%%%%%%%%%%%%%%%%%%%%%%%%%%%%%%%%%%%%%%%%%%%%%%%%%%%%%%%%%
\section{The quantitative complementarity for single kaons}\label{chapcomsingle}
%%%%%%%%%%%%%%%%%%%%%%%%%%%%%%%%%%%%%%%%%%%%%%%%%%%%%%%%%%%%%%%%
%%%%%%%%%%%%%%%%%%%%%%%%%%%%%%%%%%%%%%%%%%%%%%%%%%%%%%%%%%%%%%%%

Let us define for an arbitrary state $\tau(t)$ of a neutral kaon evolving in time the
following single partite property ($t$ is suppressed)
\begin{eqnarray}\label{singleparticlekaon}
{\cal S}^2(\tau)&=&{\cal P}^2(\tau)+{\cal V}^2(\tau)
\end{eqnarray}
where we define the predictability by
\begin{eqnarray}{\cal P}(\tau)=|Tr(\left(\begin{array}{cc}\sigma_z&0\\
0&0\end{array}\right)\tau)|=|Tr(\sigma_z\tau_{ss})|\end{eqnarray}
and the coherence by
\begin{eqnarray}{\cal V}(\tau)=2|Tr(\left(\begin{array}{cc}\sigma^+&0\\
0&0\end{array}\right)\tau)|=2|Tr(\sigma^+\tau_{ss})|\,.\end{eqnarray}
Obviously, only for the surviving components it makes sense to
obtain ``particle--like'' or ``wave--like'' information.

Moreover, different to qubits we have two different physical
options: the appearing Pauli matrices have to be defined relative to
a basis choice. In the following we assume that $\tau_{ss}$ is given
in the strangeness basis, $\{K^0,\bar K^0\}$, then the above Pauli
matrices can be defined by ``the strangeness choice''
\begin{eqnarray}\label{strangenesschoice}
S\equiv\sigma_z&:=&|K^0\rangle\langle K^0|-|\bar
K^0\rangle\langle\bar
K^0|\nonumber\\
\sigma^+&:=&|K^0\rangle\langle\bar K^0|
\end{eqnarray}
or by ``the ${\cal CP}$ choice''
\begin{eqnarray}\label{cpchoice}{\cal CP}\equiv\sigma_z&:=&|K^0_1\rangle\langle
K^0_1|-|K^0_2\rangle\langle
K^0_2|\nonumber\\
\sigma^+&:=&|K^0_1\rangle\langle K^0_2|\;.\end{eqnarray} Let us in
detail discuss these two options later.\\
\\
\textbf{Is ${\cal S}$ a useful quantity?}\\
%\\
For pure initial states $\tau$ we derive for any basis choice of the
Pauli matrices
\begin{eqnarray}
{\cal S}(\tau)&=&{\cal S}(\tau_{ss})%= {\cal S}(|\psi(t)\rangle\langle\psi(t)|)
=Tr(\tau_{ss})=\langle\psi(t)|\psi(t)\rangle\,,
\end{eqnarray}
thus one obtains the probability that a kaon survives until time
$t$, i.e. the normalization to a surviving kaon $Tr(\tau_{ss})\not
=1$ for $t>0$. Or differently stated the result is the
associated Bloch vector of the surviving block of $\tau$, i.e.
$\tau_{ss}$. This is the same result as for qubits, however, the
Bloch vector is not normalized due to the decay property. Clearly,
${\cal S}$ is basis independent.

Therefore Bohr's complementarity relation for initially pure kaon
states reads
\begin{eqnarray}
{\cal S}^2(\tau_{ss})/(Tr\,\tau_{ss})^2=1
\end{eqnarray}
which is analogous to the one for qubits, Eq.~(\ref{comp}), except
for the normalization due to decay.  If we define the dynamical
mixedness in a similar way as for qubits,
Eq.~(\ref{qubitmixedness}), ranging from $0$ to $1$
\begin{eqnarray}
M^2(\tau_{ss})=2\left((Tr\,\tau_{ss})^2-Tr(\tau_{ss}^2)\right)
\end{eqnarray}
we obtain ---in an analogous way to qubits,
Eq.~(\ref{compmixed}),--- for a general single kaon state the
following complementary relation for all states and times
\begin{eqnarray}\label{Bohrsinglekaon}
{\cal S}^2(\tau_{ss})/(Tr\,\tau_{ss})^2+
M^2(\tau_{ss})/(Tr\,\tau_{ss})^2\;=\;1\,.
\end{eqnarray}
Note that the equation solely depends on the surviving property, i.e.
$\tau_{ss}$, and cannot be re-expressed by $4\times 4$ density matrix
$\tau$ describing the full quantum state under investigation. This is also seen,
if one multiplies Eq.(\ref{Bohrsinglekaon}) by $Tr^2(\tau_{ss})$
and takes the squared root and uses
$Tr\,\tau=1=Tr\,\tau_{ss}+Tr\,\tau_{ff}$, it turns into
\begin{eqnarray}\label{completecomplement}
\sqrt{S^2(\tau_{ss})+M^2(\tau_{ss})}+Tr\,\tau_{ff}=1\;.
\end{eqnarray}
Now we can also recognize where the information flows. As the decoherence
approach of particle decay only describes the flow from the
surviving to the decaying states, $Tr\,\tau_{ff}$ is the missing
term to the information about the surviving components. Note that
this kind of information does not have the same dependence, i.e. is
not quadratically, as the information loss due to classical
uncertainty and this is also the reason why Bohr's relation cannot be re--expressed solely with $\tau$.

%Thus Eq.~(\ref{completecomplement}) is the
%complementarity relation for a single kaonic quantum system we
%searched for, containing:
%\begin{itemize}
%    \item the obtainable information $S(\tau_{ss})$, contained in the surviving part of the propagating kaonic system
%    \item the classical uncertainty $M_d(\tau_{ss})$, quantifying the missing information about the surviving
%    part of the propagating kaonic system
%    \item the information $Tr(\tau_{ff})$, that flows into the decay products
%\end{itemize}

In summary, we have shown that the defined single partite property
${\cal S}$, Eq.~(\ref{singleparticlekaon}), indeed captures the
information obtainable from the quantum system under investigation,
i.e. of a naturally interfering and decaying system, because
\begin{itemize}
\item[(1)] while predictability and coherence dependent on the
basis choice, the single partite property ${\cal S}$ is invariant to
basis transformations,
\item[(2)] for pure states ${\cal S}$ equals the normalization to
surviving kaons,
\item[(3)] ${\cal S}$ is complemented by the dynamical mixedness $M$, i.e. the the classical loss of obtainable
information due to mixing.
\end{itemize}
And therefore Eq.~(\ref{Bohrsinglekaon}) or
Eq.~(\ref{completecomplement}) is the
complementarity relation for a single kaonic quantum system we
searched for, containing:
\begin{itemize}
    \item[(1)]  the obtainable information $S(\tau_{ss})$, contained in the surviving part of the propagating kaonic system
    \item[(2)]  the classical uncertainty $M(\tau_{ss})$, quantifying the missing information about the surviving
    part of the propagating kaonic system
    \item[(3)]  the information $Tr(\tau_{ff})$, that flows into the decay products.
\end{itemize} Obviously, working with the
surviving part normalized for all times by surviving kaons gives the identical Bohr's
complementary relation, however, in the next section we show that doing that
for bipartite kaons one runs into troubles.\\
\\
\textbf{How does  $\mathcal{CP}$ violation affect
Bohr's complementary relation?}\\
\\
Clearly, the breaking of the $\mathcal{CP}$ symmetry changes the
physical states which are realized in Nature, Eq.~(\ref{states}),
however, the derived Bohr's complementary relation,
Eq.~(\ref{Bohrsinglekaon}), works for all states of single kaons,
consequently $\mathcal{CP}$ violation does not invalidate Bohr's
complementary relation. However, $\mathcal{CP}$ violation shifts the
kind of information from the predictability to the visibility and
vice versa.

\begin{figure*}
(a)\includegraphics[height=5cm, keepaspectratio=true]{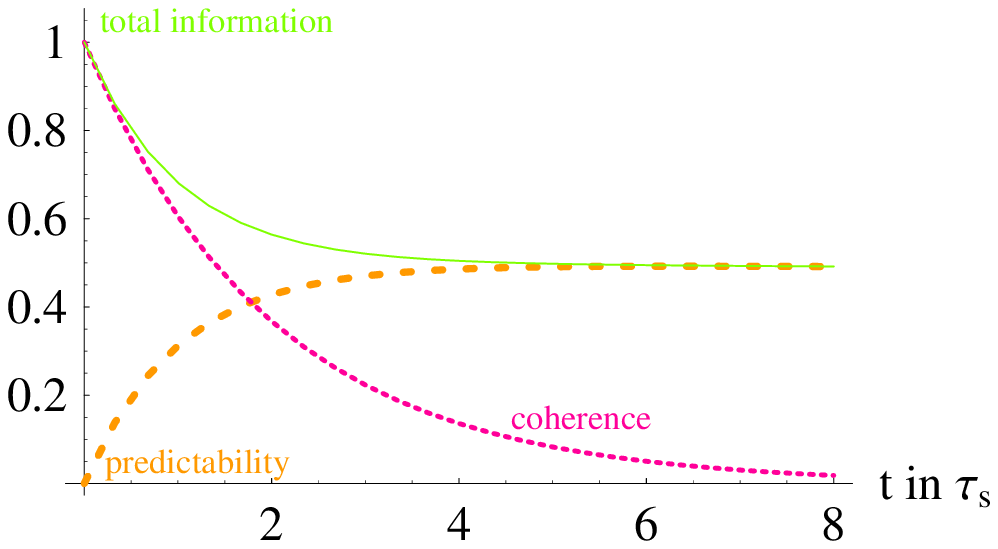}
(b)\includegraphics[height=5cm,keepaspectratio=true]{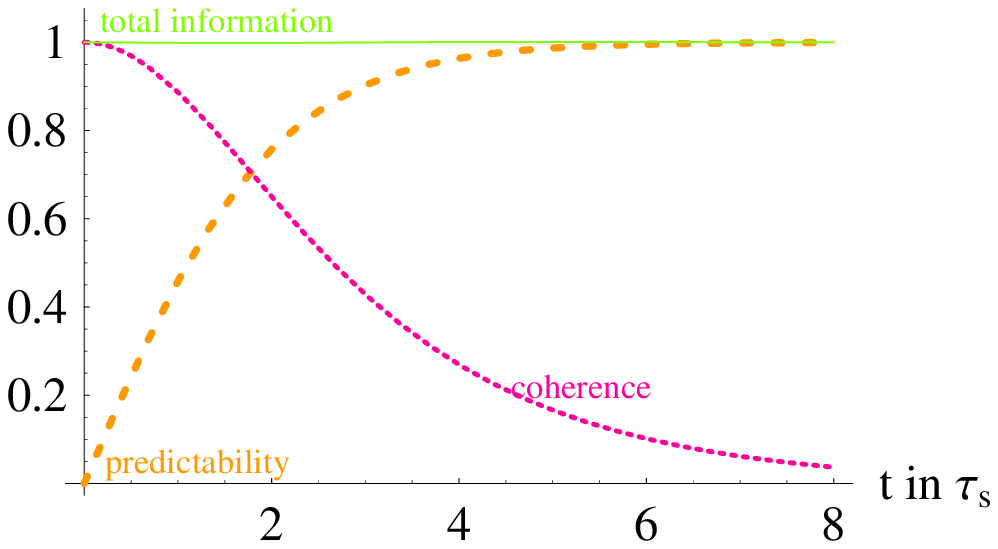}
\caption{The time evolution of the quantities, predictability ${\cal
P}$, coherence ${\cal V}$ and the total information
$Tr(\tau_{ss})^2$ are drawn for the not normalized (a) and
normalized (b) kaon state $|K^0\rangle$ in the ${\mathcal CP}$
choice (\ref{cpchoice}). At time $t=0$ the $K^0$ is an equal
superposition of the two ${\mathcal CP}$ states, i.e. the
predictability ${\cal P}$ is zero, coherence ${\cal V}$ is maximal.
As time increases the short lived state dies out and one obtains
more a priori knowledge on the ${\mathcal CP}$ eigenstates. After
about $4 \tau_S$ the system is saturated.}\label{singleK0}
\end{figure*}

Let us explicitly discuss the following examples in the ${\mathcal
CP}$ choice (\ref{cpchoice}), the time evolution of a $K_S$ and of a
$K^0$. The predictability and the coherence of a $K_S$ derives to:
\begin{eqnarray}
{\cal P}\left(|K_S(t)\rangle\langle K_S(t)|\right)&=&e^{-\Gamma_S
t}\frac{1-|\varepsilon|^2}{1+|\varepsilon|^2}\nonumber\\
{\cal V}\left(|K_S(t)\rangle\langle
K_S(t)|\right)&=&e^{-\Gamma_S t}\frac{2|\varepsilon|}{1+|\varepsilon|^2}\;.
\end{eqnarray}
Thus the a priori knowledge of being a $\mathcal{CP}$ plus or minus
state is decreased which results in a small coherence of the
non-oscillating short lived state. In Fig.~\ref{singleK0} the
predictability ${\cal P}$ and coherence ${\cal V}$ normalized and
not normalized to surviving components for a $|K^0(t)\rangle$ is
drawn.

The strangeness choice (\ref{strangenesschoice}) describes another physical situation and is discussed by the
authors of Refs.~\cite{SBGH3}. They applied Bohr's complementarity
relation in general to particle--antiparticle mixing systems,
moreover to usual quantum systems as photons, neutron,\dots but as
well to Mott scattering experiments of identical particles or
nuclei. They showed that all these two-state systems belonging to
distinct fields of physics can be treated in a unified formalism,
i.e. via the qualitative complementarity relation and moreover the
effective number of visible fringes in an experiment can be derived.

Even for specific thermodynamical quantum system Bohr's
complementary relation can be formulated and helps to understand in
a simple information theoretic way the usually complex behaviour of
such systems \cite{HV}.

%%%%%%%%%%%%%%%%%%%%%%%%%%%%%%%%%%%%%%%%%%%%%%%%%%%%%%%%%%%%%%%%
%%%%%%%%%%%%%%%%%%%%%%%%%%%%%%%%%%%%%%%%%%%%%%%%%%%%%%%%%%%%%%%%
\section{Quantitative complementarity for bipartite kaons and concurrence}
%%%%%%%%%%%%%%%%%%%%%%%%%%%%%%%%%%%%%%%%%%%%%%%%%%%%%%%%%%%%%%%%
%%%%%%%%%%%%%%%%%%%%%%%%%%%%%%%%%%%%%%%%%%%%%%%%%%%%%%%%%%%%%%%%

Two neutral kaons can also be produced in an entangled state, e.g.
at the $\Phi$-factory in Frascati kaon pairs are produced in
the spin singlet state
\begin{eqnarray}
|\psi(t=0)\rangle&=&\frac{1}{\sqrt{2}}\bigl\lbrace
|K^0\rangle\otimes |\bar K^0\rangle-|\bar K^0\rangle\otimes |
K^0\rangle\bigr\rbrace\,,
\end{eqnarray}
which is obviously analogous to the one for qubits
$|\psi\rangle=\frac{1}{\sqrt{2}}\lbrace
|0\rangle\otimes|1\rangle-|1\rangle\otimes|0\rangle\rbrace$,
however, the state for kaons evolves in time according to
Wigner--Weisskopf approximation given in Eq.~(\ref{exptime}). In
this section we analyze the Jakob--Bergou relation
(\ref{compbipartite}) for bipartite kaons.

In the previous section we have shown that for deriving the
predictability and coherence only the surviving component is
necessary. The surviving component of an arbitrary initially pure
state at time $t=0$ can be parameterized in the $\{K_S, K_L\}$ basis
in the following way
\begin{eqnarray}\label{initalbipartite}
\lefteqn{|\psi(t=0)\rangle\,=\,\frac{1}{N(0)}\biggl\lbrace}\nonumber\\
&&\;r_{SS}(0)\,|K_S\rangle\otimes|K_S\rangle+r_{SL}(0)\,|K_S\rangle\otimes|K_L\rangle\nonumber\\
&&+r_{LS}(0)\,|K_L\rangle\otimes|K_S\rangle+r_{LL}(0)\,|K_L\rangle\otimes|K_L\rangle\biggr\rbrace\,,
\end{eqnarray}
where $N(0)$ is the normalization. The time evolution of the above
surviving components are given by the Wigner--Weisskopf
approximation~(\ref{exptime}), i.e.
\begin{eqnarray}
r_{SS}(t)&=&\mathrm{r}_{SS}\, e^{i\phi_{SS}}\cdot e^{-i \lambda_S
t}e^{-i \lambda_S t}\, ,\nonumber\\
 r_{SL}(t)&=&\mathrm{r}_{SL}\, e^{i\phi_{SL}}\cdot e^{-i \lambda_S
t}e^{-i \lambda_L t}\, ,\nonumber\\
r_{LS}(t)&=&\mathrm{r}_{LS}\, e^{i\phi_{LS}}\cdot e^{-i \lambda_L
t}e^{-i \lambda_S t}\, ,\nonumber\\
r_{LL}(t)&=&\mathrm{r}_{LL}\, e^{i\phi_{LL}}\cdot e^{-i \lambda_L
t}e^{-i \lambda_L t}\, .
\end{eqnarray}
Therefore the time evolution of the surviving part of a general
bipartite kaon state is given by
\begin{eqnarray}\label{initalbipartitetime}
\lefteqn{|\psi(t)\rangle\,=\,\frac{1}{N(0)}\biggl\lbrace}\nonumber\\
&&\;r_{SS}(t)\,|K_S\rangle\otimes|K_S\rangle+r_{SL}(t)\,|K_S\rangle\otimes|K_L\rangle\nonumber\\
&&+r_{LS}(t)\,|K_L\rangle\otimes|K_S\rangle+r_{LL}(t)\,|K_L\rangle\otimes|K_L\rangle\biggr\rbrace\,.
\end{eqnarray}
Note that the above formula is not normalized for $t\geq0$ (it is divided
by $N(0)$ and not $N(t)$). Taking the partial trace gives the
surviving part of the reduced matrix, i.e. $Tr_{\neg
k}\left(|\psi(t)\rangle\langle\psi(t)|\right)$, representing the
surviving part of the states of the kaon $k$, i.e. the one moving to the left hand side or to the right hand side.

Also we want that the following equation holds for arbitrary pure
states as in the case for qubits (recall Eq.~(\ref{compbipartite}))
\begin{eqnarray}\label{compbipartitekaons}
\frac{{\cal P}^2\left(Tr_{\neg k}(\rho_{ssss})\right)+{\cal
V}^2\left(Tr_{\neg
k}(\rho_{ssss})\right)+C^2}{Tr\left(\rho_{ssss}(t)\right)^2}&=&1\nonumber\\
\end{eqnarray}
where $\rho_{ssss}(t)=|\psi(t)\rangle\langle \psi(t)|$. In this case
the quantity $C$ has to equal
\begin{eqnarray}\label{Cquantity}
C&=&(1-\delta^2)\cdot\frac{ 2 \left| r_{SS}(t) r_{LL} (t)-r_{SL}(t)
r_{LS} (t)\right|}{N(0)^2}\nonumber\\
&=&(1-\delta^2)\cdot e^{-2 \Gamma
t}\cdot\nonumber\\
&&\frac{2 \left| \mathrm{r}_{SS}\,\mathrm{r}_{LL}\,
e^{i(\phi_{SS}+\phi_{SS})}-\mathrm{r}_{SL}\,\mathrm{r}_{LS}\,
e^{i(\phi_{SL}+\phi_{LS})}\right|}{N(0)^2}\;.\nonumber\\
\end{eqnarray}
Note that the time dependence factors out and damps $C$ for any
initial state in the same way with increasing time and $C$ is in
$[0,1]$. \textbf{But is $C$ the concurrence $C(\rho(t))$ for the
states of bipartite kaons, a $16\times 16$ density matrix $\rho(t)$
and, in turn, a measure of entanglement, as it is the case for
qubits?}

\begin{theorem}
The quantity $C$, Eq.~(\ref{Cquantity}), equals the concurrence
${\cal C}(\rho_{ssss}(t))$ which is a computable function of
entanglement of formation of a $16\times 16$ density matrix
$\rho(t)=diag\{\rho_{ssss}(t),\rho_{ssff}(t),\rho_{ffss}(t),\rho_{ffff}(t)\}$
describing bipartite kaons (derived in Ref.~\cite{H2}). Here
$\rho_{ssss}(t)$ equals $|\psi(t)\rangle\langle\psi(t)|$ defined in
Eq.~(\ref{initalbipartitetime}).
\end{theorem}

\begin{proof}
In Sect.~\ref{thetimeevolutionkaon} we showed that single kaons can
be handed by an open quantum system approach, in particular by a
master equation of the Lindblad type. The density matrix $\tau(t)$
lives on $\textbf{H}_{tot}=\textbf{H}_s\bigoplus\textbf{H}_f$ where
$s$ and $f$ denote ``surviving'' and ``decaying'' or ``final''
components, and one finds for a single kaon evolving in time
\begin{equation}
\tau(t)=\left(\begin{array}{cc} \tau_{ss}(t)&0\\
0&\tau_{ff}(t)\end{array}\right)
\end{equation}
where $\tau_{ss}$ is the $2\times 2$ surviving block, i.e. is
equivalent to the time evolved state without normalization, and
$\tau_{ff}$ is $2\times 2$ block which accounts for the decaying or
final states.

Bipartite kaons are consequently described by a diagonal
$16\times16$ density matrix,
$$\rho(t)=diag\{\rho_{ssss}(t),\rho_{ssff}(t),\rho_{ffss}(t),\rho_{ffff}(t)\}$$
where the first to forth blocks refer to both kaons survive, the
kaon propagating to the left survives and the kaon propagating to the right decays, vice
versa and both kaons decay. The block $\rho_{ssss}(t)$ is equivalent
to the time evolved state Eq.~(\ref{initalbipartitetime}). For the
correctly normalized $16\times16$ state $\rho(t)$ entanglement of
formation is defined. In Ref.~\cite{H2} the author proves that it is
equivalent to the computation of concurrence of $\rho_{ssss}$ where
the complex conjugation can be done in the $\{K^0, \bar K^0\}$ basis
or in the $\{K_1, K_2\}$ basis by defining $\sigma_y$ accordingly.
Concurrence of $\rho_{ssss}$ is indeed independent of any local
basis choice and thus is the quantity we search for which completes
the single particle property of bipartite kaons. (The proof that
concurrence of $\rho_{ssss}$ is equivalent to entanglement of
formation of $\rho$ holds also for mixed initial states,
Ref.~\cite{H2}. This can also be seen by noting that $\rho$ lives on
$\textbf{H}_{tot}\otimes\textbf{H}_{tot}$, i.e.
$\textbf{H}_s\otimes\textbf{H}_s\bigoplus\textbf{H}_s\otimes\textbf{H}_f
\bigoplus\textbf{H}_f\otimes\textbf{H}_s\bigoplus\textbf{H}_f\otimes\textbf{H}_f$,
and therefore, any decomposition of $\rho$ divides into a
decomposition in each subspace. In the minimizing function of
entanglement of formation only the part in
$\textbf{H}_s\otimes\textbf{H}_s$ can contribute because there is no
entanglement between surviving and decaying parts.)
\end{proof}

\begin{figure*}
    (a)\includegraphics[height=5cm, keepaspectratio=true]{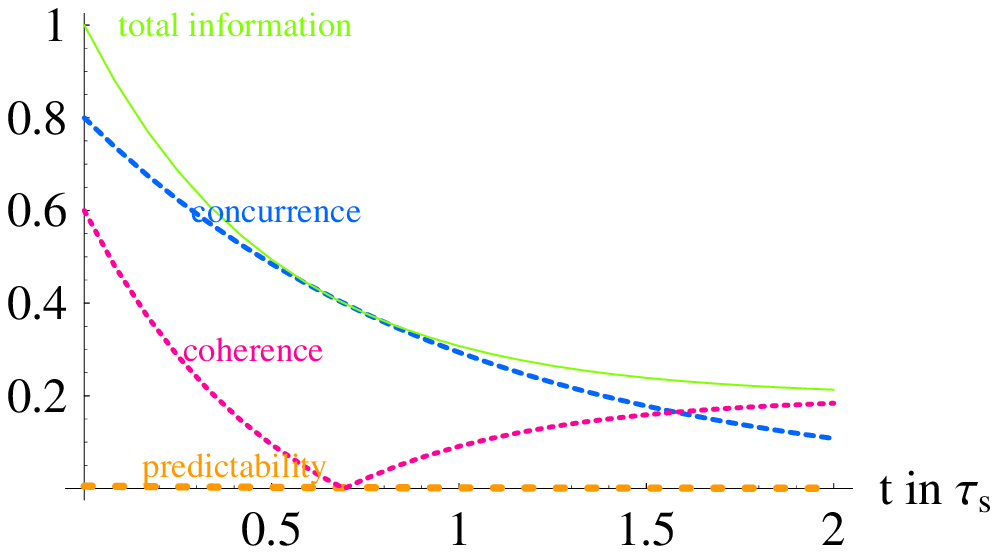}
    (b)\includegraphics[height=5cm, keepaspectratio=true]{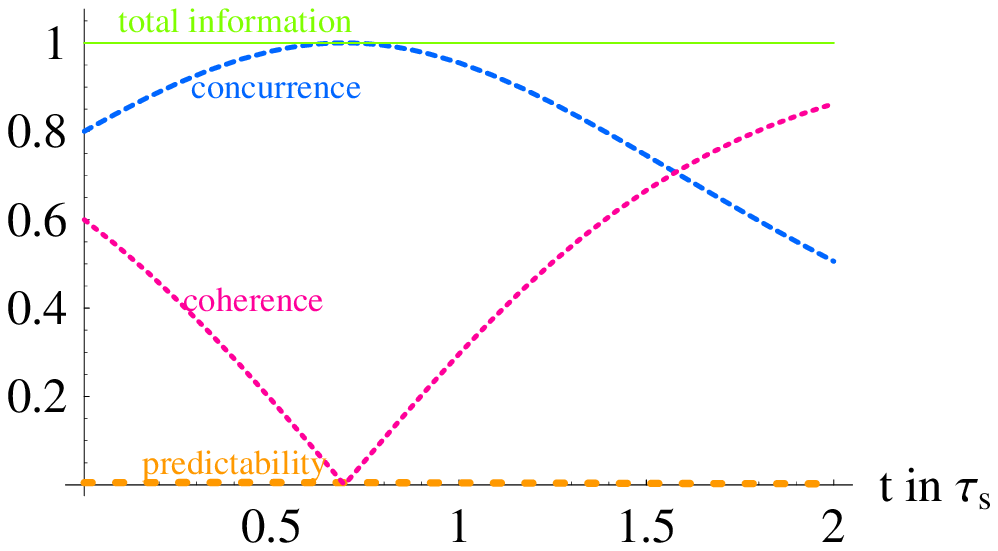}
    \caption{Predictability ${\cal
P}$, coherence ${\cal V}$, concurrence $C$ and the total information
$Tr(\rho_{ssss})^2$ are drawn for the not normalized (a) and
normalized (b) state
$|\psi\rangle=\frac{1}{\sqrt{2}}(|\psi^-\rangle_{K^0,\bar
K^0}+|K_S\rangle\otimes|K_S\rangle)$ evolving in time. Here one can
clearly see the %interpretationally disturbing
effects of normalizing to surviving pairs, concurrence rises with
time. This is due to the fact that normalizing to surviving pairs
will express the percentage contribution of the different quantities
to the total information and captures the information content of a
pair that survived until $t$, i.e. for every time point $t$ a
different state is discussed.}\label{bipartite}
\end{figure*}

\textbf{Does it make sense to work with states normalized to
surviving kaons?}\\
\\
This could prove to be quite misleading as the concurrence could
rise through the decay property. For example consider the maximally
entangled spin singlet state superposed with a fast decaying state as e.g.
$|K_S\rangle\otimes|K_S\rangle$, see also Fig.~\ref{bipartite}~(b).
Then the probability of finding a more entangled state actually does
increase with time, because normalizing to surviving kaons increases
the percentage of the entangled part. This can be understood for a
single pair, however, not in the ensemble case. If one observes for
a single pair until a certain time $t$ no decay, then the knowledge
about this pair increased, the information of the environment ``no
decay observed'' was read out. This is what is done by normalizing
to surviving pairs or differently stated it is a state preparation.
In the ensemble case the experimenter obtains the probabilities or
relative frequencies by dividing by the number of all pairs produced
at time $t=0$, not reading out the environment, but taking the
information into account which flows into the environment through
the decay property, in very same way as decoherence is described.

Thus the quantities of Fig.~\ref{bipartite}~(a) represent the
different information content of the system evolving in time,
whereas Fig.~\ref{bipartite}~(b) represent the different information
content obtained at each time point for a different state, i.e.
different experimental setups.

% We then need to acknowledge, that
%this just means that the percentage contribution of entanglement to
%the total information in our quantum system rises, while the actual
%entanglement of course decreases. And if one looks at a bunch of
%these states, that are a superposition of a more decoherence
%resistant maximally entangled state and a purely fast decaying, e.g.
%$|K_S\rangle\otimes|K_S\rangle$, then the probability of finding a
%more entangled state actually does increase with time, provided that
%one only considers the ones that have survived in both subsystems
%(See FIG.\ref{bipartite}).

But there is another interesting fact we learn. In the case of
single kaons we have noticed that due to ${\cal CP}$ violation the
quantities ${\cal P}$ and ${\cal V}$ change, however, ${\cal S}$ was
not affected.

\textbf{Is concurrence, a measure of entanglement, also affected by
$\mathcal{CP}$ violation?}\\
\\
The maximally entangled antisymmetric Bell states which is e.g.
produced in the $e^+ e^-$ machine DA$\Phi$NE is
given by
\begin{eqnarray}\label{max1}
|\psi^-\rangle_{K^0,\bar K^0}&:=&\frac{1}{\sqrt{2}}\left\lbrace
|K^0\bar K^0\rangle-|\bar K^0 K^0\rangle\right\rbrace\;=\;\nonumber\\
\frac{1}{(1-\delta^2)}|\psi^-\rangle_{K_S,
K_L}&:=&\frac{1}{\sqrt{2}\,(1-\delta^2)}\left\lbrace
|K_SK_L\rangle-|K_L K_S\rangle\right\rbrace\nonumber\\
\end{eqnarray}
The remaining maximally entangled states are given by
\begin{eqnarray}\label{max2}
|\psi^+\rangle_{K^0,\bar K^0}&=&
\frac{1}{(1-\delta^2)}(-)|\phi^-\rangle_{K_S,
K_L}\,,\nonumber\\
|\phi^+\rangle_{K^0,\bar K^0}&=& \frac{1}{(1-\delta^2)}\left\lbrace
|\phi^+\rangle_{K_S, K_L}+\delta\, |\psi^+\rangle_{K_S,
K_L}\right\rbrace\,,\nonumber\\
|\phi^-\rangle_{K^0,\bar K^0}&=&\frac{1}{(1-\delta^2)}\left\lbrace
\delta\, |\phi^+\rangle_{K_S, K_L}+|\psi^+\rangle_{K_S,
K_L}\right\rbrace\,.\nonumber\\
\end{eqnarray}
The concurrence, Eq.~(\ref{Cquantity}), derives for all
maximally entangled states to $1$ (the above set of
states are of course not the only set of maximally entangled states,
but clearly the concurrence is still maximal). This is also
seen if Eq.~(\ref{Cquantity}) is rewritten as
\begin{eqnarray}
{\cal C}(\rho_{ssss}(t))&=& {\cal
C}(\rho_{ssss}(0))\cdot(1-\delta^2)\cdot e^{-2 \Gamma t}\;.
\end{eqnarray}
The term $1-\delta^2$ compensates the ones of wave function in the lifetime basis, Eq.~(\ref{max1}) and Eq.~(\ref{max2}),
therefore concurrence is independent of $\mathcal{CP}$ violation.
Moreover, note that the time evolution factors out and the entanglement of the system at
time $t=0$ decreases for all states in the very same way, i.e. entanglement decreases due to the decay property.

%However, as the state evolves in time concurrence decreases strongly
%depending on the initial state. In the picture of a kaon as an open
%quantum system this seems obvious: The state gets due to the decay
%property more mixed, thus has to lose entanglement. For a given
%value of entanglement, e.g. measured by concurrence, the mixedness,
%$\sim(1-Tr(\rho^2))$, cannot exceed a certain value, these exposed
%states are called maximally entangled maximally mixed states (MEMS)
%\cite{MEMS}. In Ref.~\cite{H2} it was shown that kaons can exceed
%the maximal mixedness value given for qubits. This is due to the
%fact that kaons while being two--state systems are oscillating and
%decaying. Kaons obviously show a different behaviour in time. The
%symmetry breaking
%--- as clearly seen by the formulae above --- changes the weights
%and even adds new components in the basis of the exponential time
%evolution, i.e. $\lbrace K_S K_L\rbrace$. As a consequence ${\cal
%CP}$ breaking influences the time evolution of the initial state and
%due to the big difference between the two decay constants the small
%violation can get enhanced or not, strongly depending on the initial
%state.

In summary, ${\cal CP}$ violation ``deforms'' the state space of
single kaons because the time evolving states are not identical to
the  ${\cal CP}$ eigenstates and this shifts the information from
${\cal P}$ to ${\cal V}$ or vice versa, but the state space of
bipartite system is not ``deformed''. This means that the
two--particle property, concurrence, is not affected directly by
${\cal CP}$ violation. The more interesting is the result of
Ref.~\cite{BGH3} where the authors show that any non--zero $\delta$
leads to a violation of a Bell inequality, thus clearly testing a
two--particle property, namely nonlocality.

\section{Conclusions}

We have shown that applying the quantum mechanical formalism to
describe a naturally interfering decaying and ${\cal CP}$ violating
system in particle physics delivers results consistent with our
expectation. Bohr's complementarity seems to be an intrinsic feature
of our physical reality and in addition it fits very well with our
understanding of entanglement.

In detail we show that Bohr's quantitative relation can be obtained
by only operating on the surviving part of the density matrix
describing neutral kaons. Also we have demonstrated that the amount
of information decreasing through decay is equal to the probability
of obtaining a decay product, which validates the view of the kaonic
system as an open quantum system.

In the case of bipartite kaonic systems we are able to prove that
the amount of information, that should be found in an entanglement
measure according to the complementarity relation, is indeed equal
to the previously defined concurrence of the whole kaonic system.

Also new light has been shed upon the violation of a symmetry in
particle physics, by showing that it shifts obtainable information
about our reality to different aspects, without violating the
complementarity principle, i.e. from predictability ${\cal P}$ to
coherence ${\cal V}$ and vice versa.

It is also interesting that the single partite property ${\cal
S}=\sqrt{{\cal P}^2+{\cal V}^2}$ and the entanglement measure $C$, a
bipartite property, are unchanged by ${\cal CP}$ violation, whereas
in Refs.~\cite{BGH3,BH1} it is shown that any nonzero amount of
${\cal CP}$ violation leads to a violation of Bell' inequality, i.e.
testing another two--particle property, namely nonlocality. This
strengthens also the view that nonlocality and Bohr's complementing
quantity, entanglement, are different quantum features. For qudits this view was suggested
in Ref.~\cite{GisinPRL2005} and for kaons in
Ref.~\cite{H2}.

Demonstrating the working of Bohr's relation for kaons also helps to
clarify the ``old'' discussions on usual double slit scenarios, e.g.
Ref.~\cite{Luis,EnglertScully}%, as neutral
%kaons via Bohr's relation admit an interpretation as double slits
%provided freely by Nature \cite{SBGH4}, i.e. no experimental device
%as, e.g., a double slit or a beam splitter is need
. Moreover, by their time evolution the ``\textit{which width}''
information corresponding to ``\textit{which way}'' information is
changed automatically in time opening options of quantum marking and
erasure experiments \cite{SBGH1}. Because of different measurement
procedures, a special feature of kaons, the very working of a
quantum eraser can in a novel way be demonstrated \cite{SBGH6} and
an experimental realization at the DA$\Phi$NE machine (Frascati,
Rom) is under investigation.

\textbf{Acknowledgement:} Many thanks to R.A. Bertlmann
and G. Garbarino for enlightening discussions.

\end{document}